\documentclass[sigconf, dvipsnames]{acmart}

\usepackage{flushend}

\renewcommand\footnotetextcopyrightpermission[1]{} % removes footnote with conference information in first column
\usepackage{xspace}

\begin{document}

%
% The "title" command has an optional parameter, allowing the author to define a "short title" to be used in page headers.
\title{Enabling Practical Processing in and near Memory \\ for Data-Intensive Computing\vspace{-15pt}}

\author{Onur Mutlu$^a$$^,$$^b$ \quad Saugata Ghose$^b$  \quad Juan G\'{o}mez-Luna$^a$ \quad Rachata Ausavarungnirun$^b$}

\affiliation{%
  \institution{\vspace{-6pt}$^a$ETH Z{\"u}rich \qquad $^b$Carnegie Mellon University}
}

\newcommand{\changes}[0]{}
\newcommand{\changesI}[0]{}

\renewcommand{\shortauthors}{Onur Mutlu et al.}

\settopmatter{printacmref=false}

%\email{onur.mutlu@inf.ethz.ch}
%

% Modern computing systems suffer from the dichotomy between computation
% on one side and data storage/communication on the other.  Computation
% is performed only in the processor (and accelerators). All other parts
% of the system are dedicated to only storing and moving data. 
% so that computation can be done
% on it. 

%
% The abstract is a short summary of the work to be presented in the article.
\begin{abstract}

\changes{Modern computing systems suffer from the dichotomy between
  computation on one side, which is performed only in the processor
  (and accelerators), and data storage/movement on the other, which
  all other parts of the system are dedicated to.}  \changes{Due to
  this dichotomy, data moves a lot in order for the system to perform
  computation on it.}  Unfortunately, data movement is extremely
expensive in terms of energy and latency, much more so than
computation. As a result, a large fraction of system energy is spent
and performance is lost solely on moving data in a modern computing
system.

In this work, we re-examine the idea of reducing data movement by
\changes{performing Processing} in Memory (PIM). \changes{PIM places}
computation mechanisms in or near where the data is stored (i.e.,
inside the memory chips, in the logic layer of 3D-stacked logic and
DRAM, or in the memory controllers), so that data movement between the
computation units and memory is reduced or eliminated.  \changes{While
  the idea of PIM is not new, we examine two new approaches to
  enabling PIM: 1)~exploiting analog properties of DRAM to perform
  massively-parallel operations in memory, and 2)~exploiting
  3D-stacked memory technology design to provide high bandwidth to
  in-memory logic.}  \changes{We conclude by discussing work on
  solving key challenges to the practical adoption of \changesI{PIM}.}

% Even though the idea of PIM is not new, we will examine three new
% directions that have not been considered as much in the past: 1)
% performing massively-parallel bulk operations in memory by
% exploiting the analog operational properties of DRAM, with low-cost
% changes, 2) exploiting the logic layer in 3D-stacked memory
% technology in various ways to accelerate important data-intensive
% applications, \changes{and} 3) reducing or eliminating the practical
% adoption challenges against PIM with simple approaches.

\end{abstract}

%
% Keywords. The author(s) should pick words that accurately describe the work being
% presented. Separate the keywords with commas.
% \keywords{processing in memory, near data processing, memory systems, computation paradigms, DRAM, storage systems}

%
% This command processes the author and affiliation and title information and builds
% the first part of the formatted document.
\maketitle

\vspace{-0.05in}
\section{Introduction}

\sloppypar{Main memory, built using \changes{Dynamic Random Access Memory
  (DRAM),} is a major component in nearly all computing
\changes{systems, including} servers, cloud platforms, mobile/embedded
devices, and sensors.  \changes{Across these systems, the data
working} set sizes \changes{of applications} are rapidly growing, while
the need for fast analysis of such \changesI{data} % sets 
is increasing. Thus, main
memory is becoming an increasingly significant bottleneck across a
wide variety of computing systems and
applications~\cite{mutlu.imw13,mutlu.superfri15,kanev.isca15,boroumand.asplos18}.
\changes{The bottleneck has worsened in recent years, as it has become
  increasingly difficult to efficiently scale memory capacity, energy,
  cost, and performance across technology
  generations}~\cite{kang.memoryforum14,mutlu.imw13,mutlu.superfri15,mutlu2017rowhammer,kim-isca2014,salp,raidr,liu.isca13,luo.dsn14,luo.arxiv17},
as evidenced \changes{by the} RowHammer
problem~\cite{kim-isca2014,mutlu2017rowhammer,rowhammer-tcad19} in recent DRAM chips.}
% Thus, the main memory bottleneck has been worsening.

%donghyuk-ddma,lee-isca2009,rbla,yoon-taco2014,lim-isca09,
%wulf1995hitting, chang.sigmetrics16, lee.hpca13, lee.hpca15,
%chang.sigmetrics17, lee.sigmetrics17,

A major reason for the main memory bottleneck is the high energy and
latency associated with \emph{data movement}.  In today's computers,
to perform any operation on data, the processor must \changes{retrieve
  the data from main memory}.  \changes{This requires the memory
  controller to issue} commands to \changes{a DRAM module} across a
relatively slow and power-hungry off-chip bus (known as the
\emph{memory channel}).  The DRAM module \changes{sends the requested
  data across} the memory channel, after which the data is placed
\changes{in the caches} and registers.  The CPU can perform
computation on the data once the data is in its registers.  Data
movement from the DRAM to the CPU incurs long latency and consumes
\changes{significant} energy~\cite{hashemi.isca16,cont-runahead,
  ahn.tesseract.isca15,ahn.pei.isca15,boroumand.asplos18}.  These
costs are often exacerbated by the fact that much of the data brought
into the caches is \emph{not reused} by the
CPU~\cite{qureshi.isca07,qureshi-hpca07}, providing little benefit in
return for the high latency and energy cost.

The cost of data movement is a fundamental issue with the
\emph{processor-centric} nature of contemporary computer systems.  The
CPU is considered the master in the system, and computation is
performed only in the processor (and accelerators). In contrast, data
storage and communication units, including the main memory, are
treated as unintelligent workers that are incapable of computation. As
a result of this processor-centric design paradigm, data moves a lot
in the system between the computation units and communication/storage
units so that computation can be done on it. With the increasingly
\emph{data-centric} nature of contemporary and emerging applications,
the processor-centric design paradigm leads to great inefficiency in
performance, energy and \changes{cost: for example,
most of the real estate within a single compute node is
already dedicated to handling data movement and storage
(e.g., large caches, memory controllers, interconnects, and
main memory), and our recent work shows that 62\% of
the entire system energy of a mobile device is spent on data movement
between the processor and the memory hierarchy for widely-used
mobile workloads~\cite{boroumand.asplos18}.}

The huge overhead of data movement in modern systems along with
\changes{technology advances that enable better integration of} memory
and logic have recently prompted the re-examination of an old idea
\changes{that we will generally call \emph{Processing in Memory}
  (PIM).}  The key idea is to place computation mechanisms in or near
where the data is stored (i.e., inside the memory chips, in the logic
layer of \changes{3D-stacked DRAM}, in the memory controllers, or
inside large caches), so that data movement between where the
computation is done and where the data is stored is reduced or
eliminated, compared to \changes{contemporary} processor-centric
systems.

%\footnote{We treat the terms \emph{near-data processing} (NDP) and PIM
%  interchangeably in this work.}

The idea of PIM has been around for at least four
decades~\cite{stone1970logic,shaw1981non,elliott1992computational,kogge1994execube,
  gokhale1995processing,patterson1997case,oskin1998active,kang2012flexram,Draper:2002:ADP:514191.514197,Mai:2000:SMM:339647.339673,elliott.dt99,riedel.1998,keeton.1998,kaxiras.1997,acharya.1998}.
However, past efforts were \emph{not} \changes{widely adopted for}
various reasons, including 1)~the difficulty of integrating processing
elements with DRAM, 2)~the \changes{lack of critical memory-related
  scaling challenges that current technology and applications face}
today, and~3) that the data movement bottleneck was not as critical to
system cost, energy and performance as it is today.  \changes{We
  believe it is crucial to re-examine PIM} today with a fresh
perspective (i.e., with novel approaches and ideas), by exploiting new
memory technologies, with realistic workloads and systems, and with a
mindset to ease adoption and feasibility.

%As a result of advances in modern memory architectures, e.g., the
%integration of logic and memory in a 3D-stacked manner, various
%recent works explore a range of PIM architectures for multiple
%different purposes (e.g.,~\cite{zhu2013accelerating,
%pugsley2014ndc,zhang.hpdc14,farmahini-farahani.hpca15,ahn.tesseract.isca15,ahn.pei.isca15,loh2013processing,hsieh.isca16,pattnaik.pact16,DBLP:conf/isca/AkinFH15,impica,DBLP:conf/sigmod/BabarinsaI15,DBLP:conf/IEEEpact/LeeSK15,DBLP:conf/hpca/GaoK16,chi.isca16,gu.isca16,kim.isca16,asghari-moghaddam.micro16,boroumand2016pim,hashemi.isca16,cont-runahead,GS-DRAM,liu-spaa17,gao.pact15,guo2014wondp,sura.cf15,morad.taco15,hassan.memsys15,li.dac16,kang.icassp14,aga.hpca17,shafiee.isca16,seshadri2013rowclone,Seshadri:2015:ANDOR,chang.hpca16,seshadri.arxiv16,seshadri.micro17,nai2017graphpim,kim.arxiv17,kim.bmc18,li.micro17,kim.sc17,boroumand.asplos18}).

In \changes{this paper}, we explore two new approaches to enabling
\changes{PIM} in modern systems.  The first \changes{approach only}
\emph{minimally changes memory chips} to perform simple yet powerful
common operations that the chip \changesI{is inherently efficient at
performing~\cite{seshadri.bookchapter17,seshadri2013rowclone,chang.hpca16,kevinchang-thesis,seshadri.thesis16,Seshadri:2015:ANDOR,seshadri.arxiv16,seshadri.micro17,li.micro17,GS-DRAM,ghose.bookchapter19,ghose.bookchapter19.arxiv,deng.dac2018,mutlu2019micpro}.}
\changes{Such solutions} take advantage of the existing memory
\changes{design to perform} \emph{bulk operations} (i.e., operations
on an entire row of DRAM cells), such as bulk copy, data
initialization, and bitwise
operations~\cite{seshadri2013rowclone,chang.hpca16,Seshadri:2015:ANDOR,seshadri.arxiv16,seshadri.micro17}.
The second approach \changes{enables \changesI{PIM} in a more general-purpose
  manner by taking advantage of \changesI{emerging \emph{3D-stacked memory
    technologies}}~\cite{loh2013processing,pugsley2014ndc,zhu2013accelerating,DBLP:conf/isca/AkinFH15,impica,ahn.tesseract.isca15,nai2017graphpim,DBLP:conf/sigmod/BabarinsaI15,gao.pact15,kim.bmc18,gu.isca16,boroumand2016pim,boroumand.isca19,boroumand.asplos18,chi.isca16,kim.isca16,DBLP:conf/IEEEpact/LeeSK15,ahn.pei.isca15,zhang.hpdc14,hsieh.isca16,pattnaik.pact16,sura.cf15,hassan.memsys15,farmahini-farahani.hpca15,DBLP:conf/hpca/GaoK16,guo2014wondp,liu-spaa17,kim.sc17}.}
\changes{3D-stacked memory chips have much greater \emph{internal}
  bandwidth than is available externally on the memory
  channel~\cite{lee.taco16}, and many such chip architectures (e.g.,
  Hybrid Memory Cube~\cite{hmc.spec.1.1,hmc.spec.2.0}, High-Bandwidth
  Memory~\cite{jedec.hbm.spec,lee.taco16}) include a \emph{logic
    layer} where designers can add some processing logic (e.g.,
  accelerators, simple cores, reconfigurable logic) that can take
  advantage of this high internal bandwidth.}

% In order to stack multiple layers of memory, 3D-stacked chips use
% vertical \emph{through-silicon vias} (TSVs) to connect the layers to
% each other, and to the I/O drivers of the chip~\cite{lee.taco16}.  The
% TSVs provide much greater \emph{internal} bandwidth than is available
% externally on the memory channel.  Several such 3D-stacked memory
% architectures, such as the Hybrid Memory
% Cube~\cite{hmc.spec.1.1,hmc.spec.2.0} and High-Bandwidth
% Memory~\cite{jedec.hbm.spec,lee.taco16}, include a \emph{logic layer},
% where designers can add some processing logic (e.g., accelerators,
% simple cores, reconfigurable logic) to take advantage of the high
% internal bandwidth.
%%% ONUR: Maybe cut the above part starting from ``In order to''...

Regardless of the approach taken to PIM, there are key practical
adoption challenges that system architects and programmers must
address to enable the widespread adoption of \changesI{PIM} across the computing
landscape and in different domains of workloads. We also briefly
discuss these challenges in this paper, along with references to some
existing work that addresses these challenges.

\vspace{-0.05in}
\section{Minimally Changing Memory Chips}
\label{sec:minimally}

\changes{Minimal modifications in existing memory chips can} enable
simple yet powerful computation capability inside the chip.
\changes{These modifications take} advantage of the existing
interconnects in and analog operational behavior of conventional
memory chips, e.g., DRAM \changes{architectures, without} the need for
a logic layer and usually without the need for logic processing
elements. As a result, the overheads imposed on the memory chip are
low.  \changes{There are a number of mechanisms that use this approach
  to take advantage of the high internal bandwidth available within
  each memory cell array}~\cite{seshadri.bookchapter17,seshadri2013rowclone,chang.hpca16,kevinchang-thesis,seshadri.thesis16,Seshadri:2015:ANDOR,
  seshadri.arxiv16, seshadri.micro17}.  We briefly describe one such
design, Ambit, which enables in-DRAM bulk bitwise
operations~\cite{Seshadri:2015:ANDOR,
  seshadri.arxiv16,seshadri.micro17}, by building on RowClone, which
enables fast and energy-efficient in-DRAM data
movement~\cite{seshadri2013rowclone,chang.hpca16}.

%  Then, we describe a low-cost substrate that
%  {performs} data reorganization for non-unit strided access
%  patterns~\cite{GS-DRAM}.

%RowClone, which enables in-DRAM
%bulk data movement operations~\cite{seshadri2013rowclone} and 

%\subsection{Ambit: In-DRAM Bulk Bitwise Operations}
%\label{sec:ambit}

%~\cite{bmide,bmidc,fastbit,bicompression}
%~\cite{li.dac16}
%enc1
%bitwise-alignment,

%\vspace{0.05in}
{\bf Ambit: In-DRAM Bulk Bitwise Operations.} Many applications use
\changes{\emph{bulk bitwise
    operations}~\cite{btt-knuth,hacker-delight} (i.e., bitwise
  operations on large bit vectors), such as} bitmap indices, bitwise
scan acceleration~\cite{bitweaving} for databases, accelerated
document filtering for web search~\cite{bitfunnel}, DNA sequence
\changesI{alignment~\cite{xin.shd.bioinformatics15,
  alser.bioinformatics17,shouji-bioinformatics19,kim.bmc18}, encryption} algorithms~\cite{xor1,xor2}, graph processing, and
networking~\cite{hacker-delight}. Accelerating bulk bitwise operations
can {thus} significantly boost the performance and energy efficiency
of a wide range of applications.

\changes{We} have recently proposed a new
\textbf{A}ccelerator-in-\textbf{M}emory for bulk \textbf{Bit}wise
operations (Ambit)~\cite{Seshadri:2015:ANDOR, seshadri.arxiv16,
  seshadri.micro17}.  Unlike prior approaches, Ambit uses the analog
operation of existing DRAM technology to perform bulk bitwise
operations.  Ambit has two components.  The first component,
Ambit--AND--OR, implements a new operation called \emph{triple-row
  activation}, where the memory controller simultaneously activates
three rows.  \changes{Triple-row activation uses the charge sharing
  principles that govern the operation of the DRAM array to perform a
  bitwise AND or OR on two rows of data, by controlling the initial
  value on the third row.}
% Triple-row activation performs a bitwise majority
% function across the cells in the three rows, due to the charge sharing
% principles that govern the operation of the DRAM array.  By
% controlling the initial value of one of the three rows, we can use
% triple-row activation to perform a bitwise AND or OR of the other two
% rows.  
The second component, Ambit--NOT, takes advantage of the two inverters
that are connected to each sense amplifier in a DRAM subarray,
\changes{as the voltage level of one of the inverters represents the
  negated logical value of the cell.}  The Ambit design adds a special
row to the DRAM array \changes{to capture this negated value.}  One
possible implementation of the special row~\cite{seshadri.micro17} is
a row of \emph{dual-contact cells} (a 2-transistor 1-capacitor
cell~\cite{2t-1c-1,migration-cell}), each connected to both inverters
inside a sense amplifier.  \changes{Even in the presence of process
  variation (see \cite{seshadri.micro17}), Ambit can reliably perform
  AND, OR, and NOT operations completely using DRAM technology, making
  it functionally (i.e., Boolean logic) complete.}
% With the ability to perform AND, OR, and
% NOT operations \changes{completely using DRAM technology}, 
% Ambit is functionally (i.e., Boolean logic) \changes{complete.}
% % : \changes{it} can reliably perform \emph{any} bulk bitwise operation completely
% % using DRAM technology within a DRAM chip, 
% even} in the presence of
% process variation (see \cite{seshadri.micro17} for details).

Ambit provides promising performance and energy improvements. Averaged
across seven commonly-used bulk bitwise operations (NOT, AND, OR,
NAND, NOR, XOR, XNOR), Ambit with 8 DRAM banks improves bulk bitwise
operation throughput by 44$\times$ compared to an Intel Skylake
processor~\cite{intel-skylake}, and 32$\times$ compared to the NVIDIA
GTX 745 GPU~\cite{gtx745}. Compared to DDR3 DRAM, Ambit reduces energy
consumption by 35$\times$ on \changes{average. 
% Compared to performing the operations in the logic layer of the HMC
% 2.0~\cite{hmc.spec.2.0} 3D-stacked memory, Ambit improves bulk
% bitwise operation throughput by 2.4$\times$.
When} integrated directly into the HMC 2.0 device, which has many more
banks, Ambit improves operation throughput by 9.7$\times$ compared to
processing in the logic layer of HMC 2.0.  \changes{Our work evaluates
  the end-to-end benefits of Ambit on real database queries using
  Bitmap indices and the BitWeaving database~\cite{bitweaving},
  showing query latency reductions of 2X to 12X, with larger benefits
  for larger data set sizes.}
% Our work evaluates end-to-end benefits of Ambit on real database
% queries using Bitmap indices as well as the BitWeaving
% database~\cite{bitweaving}: we find that Ambit reduces query latency
% by 2X to 12X, providing larger benefits when data set sizes are
% larger.

A number of Ambit-like bitwise operation substrates have been proposed
in recent years, making use of emerging resistive memory technologies,
e.g., phase-change memory (PCM)~\cite{lee-isca2009,lee.ieeemicro10,
  lee.cacm10, zhou.isca09,qureshi.isca09,yoon-taco2014}, SRAM, or
specialized DRAM. These substrates can perform bulk bitwise operations
in a special DRAM array augmented with computational
circuitry~\cite{li.micro17} and in PCM~\cite{li.dac16}. Similar
substrates can perform simple arithmetic operations in
SRAM~\cite{aga.hpca17,kang.icassp14} and arithmetic and logical
operations in memristors~\cite{kvatinsky.tcasii14, kvatinsky.tvlsi14,
  shafiee.isca16}.
% We believe it is extremely important to continue
% exploring such low-cost Ambit-like substrates, as well as more
% sophisticated computational substrates, for all types of memory
% technologies, old and new. 
% Resistive memory technologies are
% fundamentally non-volatile and amenable to in-place updates, and as
% such, can lead to even less data movement compared to DRAM, which
% fundamentally requires some data movement to read any data. Thus, we
% believe it is very promising to examine the design of emerging
% resistive memory chips that can incorporate Ambit-like bitwise
% operations and other types of suitable computation capability.
\changes{Resistive memory technologies are amenable to
in-place updates, and can thus incorporate Ambit-like operations
with even less data movement than DRAM.
Thus, we believe it is extremely important to continue
exploring low-cost Ambit-like substrates, as well as more
sophisticated computational substrates, for all types of memory
technologies, old and new.}
% , making them a
% very promising substrate for bulk in-memory operations.

\vspace{-0.05in}
\section{PIM using 3D-Stacked Memory}
\label{sec:3dstacked}

Several works propose to place some form of processing logic
(typically accelerators, simple cores, or reconfigurable logic) inside
the logic layer of 3D-stacked memory~\cite{lee.taco16}.  This
\emph{\changesI{PIM} processing logic}, which we also refer to as \emph{\changesI{PIM}
  cores}, can execute portions of applications (from individual
instructions to functions) or entire threads and applications,
depending on the design of the architecture. 
\changes{The \changesI{PIM} cores connect to the memory stacks that are on
top of them using vertical \emph{through-silicon vias}~\cite{lee.taco16}, 
which provide high-bandwidth and low-latency access to data.}
% Such PIM cores have high-bandwidth and low-latency access to the
% memory stacks that are on top of , since the logic layer and the
% memory layers are connected via high-bandwidth vertical connections
% the vertical through-silicon vias~\cite{lee.taco16}, e.g.,
% through-silicon vias.
In this section, we discuss examples of how systems can make use of
relatively simple \changesI{PIM \changes{cores to}} avoid data movement and thus
obtain significant performance and energy improvements for a variety
of application domains.

%\subsection{Tesseract: Graph Processing}
%\label{sec:tesseract}

%~\cite{salihoglu.ssdbm13, tian.vldb13, low.vldb12,
%  hong.asplos12, malewicz.sigmod10, harshvardhan.pact14,
%  gonzalez.osdi12, ligra, Seraph, graphlab, nai2017graphpim}

%\vspace{0.05in}
{\bf Tesseract: Graph Processing.} A popular modern application is
large-scale graph processing/analytics.  Graph processing has broad
applicability and use in many domains, from social networks to machine
learning, from data analytics to bioinformatics.  Graph analysis
workloads put large pressure on memory bandwidth due to 1) frequent
random memory accesses across large memory regions (leading to limited
cache efficiency and unnecessary data transfer on the memory bus) and
2) small amount of computation per data item fetched from memory
(leading to limited ability to hide long memory latencies and
exercising the memory energy bottleneck).  These two characteristics
make it very challenging to scale up such workloads despite their
inherent parallelism, especially with conventional architectures based
on large on-chip caches and relatively scarce off-chip memory
bandwidth for random access.
%%% ONUR: shorten the above paragraph...

% We exploit the high bandwidth and the computation capability
% available within the logic layer of 3D-stacked memory to overcome
% the limitations of conventional architectures for graph processing.
% To this end,
\changes{To overcome the limitations of conventional architectures,}
we design Tesseract, a programmable PIM accelerator for large-scale
graph processing~\cite{ahn.tesseract.isca15}.  Tesseract consists of
1) \changes{simple in-order \changesI{PIM} cores that exploit the high memory
  bandwidth available in the logic layer of 3D-stacked memory, where
  each core manipulates data only on the memory partition it is
  assigned to} control, 2) an efficient \changes{communication
  interface that allows a \changesI{PIM} core} to request computation on data
elements that reside in the memory partition controlled by another
core, and 3) a message-passing based programming interface, similar to
how modern distributed systems are programmed, which enables remote
function calls on data that resides in each memory partition.
Tesseract moves functions to data rather than moving data elements
across different memory partitions and \changes{cores.
% It also includes two hardware prefetchers specialized for memory
% access patterns of graph processing.
Our} comprehensive evaluations using five state-of-the-art graph
processing workloads with large graphs show that \changes{Tesseract}
improves average system performance by {$13.8\times$} and reduces
average system energy by 87\% over a state-of-the-art conventional
system.

%\subsubsection{Google Consumer Workloads}
%\subsection{Consumer Workloads}
\label{sec:google}

{\bf Consumer Workloads.} A popular domain of computing is consumer
devices, including smartphones, tablets, web-based computers
\changes{(e.g., Chromebooks)}, and wearable devices.  In such devices,
energy efficiency is a first-class concern due to the limited battery
capacity and the stringent thermal power budget.  We find that
\emph{data movement} is a major contributor to \changes{energy} (and
execution time) in modern consumer devices: across four popular
\changes{workloads (described} next), \changes{62.7\%} of the total
system energy, on average, is spent on data movement across the memory
hierarchy~\cite{boroumand.asplos18}.

%~\cite{chrome}
%~\cite{mobile-tensorflow}
%~\cite{vp9-specification}

% , which account for a significant
% portion of the applications executed on consumer devices.  These
% workloads include (1)~the Chrome web browser, a popular browser used
% in mobile devices and laptops; (2)~TensorFlow Mobile, Google's machine
% learning framework, which is used in services such as Google
% Translate, Google Now, and Google Photos; (3)~the VP9 video playback
% engine, and (4)~the VP9 video capture engine, both of which are used
% in many video services such as YouTube and Google Hangouts.  

We comprehensively analyze the energy and performance impact of data
movement for several widely-used Google consumer
workloads~\cite{boroumand.asplos18}:
\changes{1)~the Chrome web browser,
2)~TensorFlow Mobile (Google's machine learning framework),
3)~the VP9 video playback engine, and
4)~the VP9 video capture engine.}
We find that offloading key functions (called {\em target functions})
of these workloads to \changes{\changesI{PIM} logic}
greatly reduces data movement.  \changes{However, 
consumer} devices are extremely stringent in terms
of the extra area and energy they can accommodate.  As a result, it is
important to identify what kind of PIM logic can both 1) maximize
energy efficiency and 2)~be implemented at minimum possible 
\changes{area and energy costs.}

We find that many of the target functions for PIM in consumer
workloads are comprised of simple operations \changes{(e.g.,
  \emph{memcopy}/\emph{memset}, basic arithmetic and bitwise
  operations), and can be implemented easily in the logic layer using
  either 1)~a small low-power general-purpose core or 2)~small
  fixed-function accelerators.}
% , such as \emph{memcopy},
% \emph{memset}, basic arithmetic and bitwise operations, and simple
% data shuffling and reorganization routines.  Therefore, we can
% relatively easily implement \changes{these target} functions in the logic
% layer of 3D-stacked memory using either 1)~a small low-power
% general-purpose embedded core or 2)~a group of small fixed-function
% accelerators.  
Our analysis shows that the area of a PIM core and a
PIM accelerator take up no more than 9.4\% and 35.4\%, respectively,
of the area available for \changesI{PIM} logic in an HMC-like~\cite{hmc.spec.2.0}
3D-stacked memory architecture.  Both the PIM core and PIM accelerator
eliminate a large amount of data movement, and thereby significantly
reduce total system energy (by an average of 55.4\% across {all} the
workloads) and execution time (by an average of 54.2\%).

\vspace{-0.05in}
\section{Enabling PIM Adoption}
\label{sec:adoption}

% We briefly touch upon key challenges that must be addressed for PIM
% to be adopted as a mainstream architecture in a wide variety of
% systems and workloads, and in a seamless manner that does not place
% heavy burden on the vast majority of programmers.

% (e.g., by appropriate runtime scheduling of code to PIM logic and
% mapping of data so that data movement is minimized and
% Quality-of-Service demands of applications are
% satisfied)

Pushing computation from the CPU into memory introduces new challenges
for system architects and programmers to overcome.  \changes{Many of
  these challenges must be addressed for \changesI{PIM} to be adopted in a wide
  variety of systems of workloads, without placing a heavy burden on
  most \changesI{programmers~\cite{mutlu2019micpro,ghose.bookchapter19.arxiv}}}  These challenges include 1) how to easily
program PIM systems (with good programming model, library, compiler
and tools support)~\cite{ahn.pei.isca15,hsieh.isca16}; 2) how to
design runtime systems and system software that can take advantage of
PIM \changes{(e.g., runtime scheduling of code on \changesI{PIM} logic, data
  mapping)}~\cite{ahn.pei.isca15,hsieh.isca16,pattnaik.pact16,
  boroumand.asplos18}; 3) how to efficiently enable coherence between
PIM logic and CPU/accelerator cores that operate on \changesI{shared
data~\cite{boroumand2016pim,boroumand.isca19,ahn.pei.isca15}; 4)} how
to efficiently enable virtual memory support on the PIM
logic~\cite{impica}; 5) how to design high-performance data structures
for PIM whose performance is better than concurrent data structures on
multi-core machines~\cite{liu-spaa17}; 6) how to accurately assess the
benefits and shortcomings of \changesI{PIM} using realistic workload suites,
rigorous analysis methodologies, and accurate and flexible simulation
infrastructures~\cite{ramulator, ramulator.github}.

We believe these challenges provide exciting cross-layer research
opportunities. Fundamentally solving the data movement problem
requires a paradigm shift to a data-centric computing system design,
where computation happens in or near memory, with minimal data
movement. We argue that research enabled towards such a paradigm shift
would be very useful for both PIM as well as other potential ideas
that can reduce data movement.

\vspace{-0.1in}
\section*{Acknowledgments}
\changesI{
We thank members of the SAFARI Research Group and collaborators at Carnegie Mellon, ETH Zurich, and other universities, who have contributed to the various works we describe in this paper. Thanks also to our research group's industrial sponsors over the past ten years, especially Alibaba, Google, Huawei, Intel, Microsoft, NVIDIA, Samsung, and VMware. This work was also partially supported by the Semiconductor Research Corporation and NSF.
}
\vspace{-0.1in}
     
% The next two lines define the bibliography style to be used, and the bibliography file.
\bibliographystyle{ACM-Reference-Format}
\bibliography{sample-sigconf}

\end{document}